\documentclass[showpacs,preprintnumbers,superscriptaddress]{revtex4}
\usepackage{CJK}
\usepackage{amsmath,amssymb,graphicx,bm}
\begin{document}

\title{Exact solution for accretion onto a moving charged dilaton black hole}

\author{Rong-Jia Yang \footnote{Corresponding author}}
\email{yangrongjia@tsinghua.org.cn}
\affiliation{College of Physical Science and Technology, Hebei University, Baoding 071002, China}
\affiliation{Hebei Key Lab of Optic-Electronic Information and Materials, Hebei University, Baoding 071002, China}
\affiliation{National-Local Joint Engineering Laboratory of New Energy Photoelectric Devices, Hebei University, Baoding 071002, China}
\affiliation{Key Laboratory of High-pricision Computation and Application of Quantum Field Theory of Hebei Province, Hebei University, Baoding 071002, China}

\author{Yinan Jia}
\affiliation{College of Physical Science and Technology, Hebei University, Baoding 071002, China}
\author{Lei Jiao}
\affiliation{College of Physical Science and Technology, Hebei University, Baoding 071002, China}
\affiliation{School of Physics and Astronomy, Sun Yat-Sen University, Zhuhai 519082, China}

\begin{abstract}
We present an analytic solution for accretion of a gaseous medium with adiabatic equation of state onto a charged dilaton black hole which moves at a constant velocity. We determine the four-velocity of accreted flow and find that it possesses axial symmetry. We obtain the particle number density and the accretion rate which depend on the mass, the magnetic charge, and the dilation of black hole, meaning that these parameters take important roles in the process of accretion. Possible theoretical and observational constraints on the parameter related to the dilation are discussed. The results may help us to get deeper understanding of the behavior of accreted flow near the event horizon of black hole.
\end{abstract}

\maketitle

\section{Introduction}
Accretion of matter onto a massive object is a long-standing interesting phenomenon
in astrophysics \cite{Yuan:2014gma}. The pioneers works in this field were published in \cite{hoyle1939effect, 1940Obs6339L, Bondi:1944jm, Bondi:1952ni, michel1972accretion}. Since then accretion has been an extensively studied subject in the literature, see for example \cite{begelman1978accretion, Malec:1999dd, Babichev:2004yx, Babichev:2010kj, Rodrigues:2016uor, Contreras:2018gct, Abbas:2018ygc, Zheng:2019mem, Yang:2020bpj, UmarFarooq:2020aum, Nozari:2020swx, Panotopoulos:2021ezt, Iftikhar:2020ykp, Gao:2008jv, John:2013bqa, Jiao:2016iwp, Ganguly:2014cqa, Mach:2013fsa, Kremer:2020yfg, Tejeda:2019lie, Feng:2022bst}. These studies mainly focused on accretion rate, critical radius, flow parameters, and so on. In fact, we can also use accretion to open up new possibilities. In \cite{Yang:2015sfa}, attempts were made to test the asymptotically safe scenario via the quantum correction to accretion onto a renormalization-group-improved Schwarzschild black hole; while in \cite{Yang:2018zef} accretion onto a Schwarzschild-like black hole was considered to constrain the parameter characterizing the breaking of Lorentz symmetry. It was shown in \cite{Jamil:2008bc, Yang:2019qru} that the conditions under which the accretion is possible give limits on the ratio of mass to charge. Numerical analysis, in general, are needed to dealt with nonspherical accretion for either Newtonian or relativistic flow \cite{Papadopoulos:1998up, Font:1998sc, Zanotti:2011mb, Lora-Clavijo:2015hqa}. So finding analytical solutions is often a very valuable task. An exact fully relativistic solution describing stationary accretion of matter with adiabatic equation of state onto a moving Schwarzschild or Kerr black hole was derived in \cite{Petrich:1988zz}. This work was generalized to the Kerr-Newman metric case \cite{Babichev:2008dy}. An analytic nonspherical solution was also obtained for accretion onto a moving Reissner-Nordstr\"{o}m black hole \cite{Jiao:2016uiv}. In \cite{Liu:2009ts, Zhao:2018ani}, analytic solutions were presented for dust shells onto a Schwarzschild black hole. Recently, an exact solution representing accretion of collisionless Vlasov gas onto a moving Schwarzschild black hole was given in \cite{Mach:2021zqe, Mach:2020wtm} which based on the Hamiltonian formalism developed in \cite{Rioseco:2016jwc} which allows for an analysis of more complex flows on the fixed Schwarzschild background.

Exact solutions are important to accurately and directly describe the behavior of accreted matters near a black hole. However, in general, the analytical solution for accreting process is very complex and difficult to obtain. So it is valuable to find analytical solutions for accretion onto different black holes. Here we aim to analysis the accreting process for a moving charged dilaton black hole which is a solution to low-energy string theory representing a static, spherically symmetric charged black hole \cite{Gibbons:1987ps, Garfinkle:1990qj}. The results obtained here and in the literatures about analytic solutions may be helpful to understand the physical mechanism of accretion onto black holes.

In the following Section II, we briefly review the fundamental equations for accretion onto a moving black hole. In Section III, we hope to derive an exact solution for accretion onto a moving charged dilaton black hole. Finally, we will briefly summarize our results in section IV.

\section{Fundamental equations}
In this section, we briefly review some basic equations in the process of accretion. We adopt the reference frame in which the black hole is rest while the homogeneous medium moves at a constant velocity at infinity. Assuming the flow of matter is a perfect fluid, then the relativistic vorictity tensor is given by \cite{Petrich:1988zz}
\begin{eqnarray}
\label{1}
\omega_{\mu\nu}=P^{\alpha}_{\mu}P^{\beta}_{\nu}[(hu_{\alpha})_{;\beta}-(hu_{\beta})_{;\alpha}],
\end{eqnarray}
where $u^{\mu}$ is the component of four-velocity, $h\equiv(\rho+P)/n$ is the enthalpy, and $P_{\mu}^{\nu}=\delta_{\mu}^{\nu}+u_{\mu}u^{\nu}$ is the projection tensor, and the semicolon denotes the covariant derivative with respect to the coordinate. Throughout this paper, we use the units including $c=G=1$, where $c$ and $G$ are the speed of light and Newtonian gravitational constant, respectively. We take the signs of components of metric tensor of Minkowski space-time as $(-, +, +, +)$. Since the fluid is perfect, Euler's equation becomes
\begin{eqnarray}
\label{2}
(hu_{\mu})_{;\alpha}u^{\alpha}+h_{,\mu}=0,
\end{eqnarray}
where the comma denotes the ordinary derivative with respect to the coordinate. From the Eqs. (\ref{1}) and (\ref{2}), one can obtain a simple expression for the vorticity
\begin{eqnarray}
\label{3}
\omega_{\mu\nu}=[(hu_{\mu})_{;\nu}-(hu_{\nu})_{;\mu}].
\end{eqnarray}
The constant velocity at which the black hole moves and the homogeneity of medium upstream imply that the vorticity is zero. And if the vorticity is zero on some initial hypersurface, it will be zero everywhere, like the case in Newtonian flow. Thus the quantity $hu_{\mu}$ can be expressed as the gradient of a potential \cite{1980ApJ...235.1038M}:
\begin{eqnarray}
\label{4}
hu_{\mu}=\psi_{,\mu}.
\end{eqnarray}
If there no particles are destroyed or created, the equation of continuity for particle density $n$ can be written as
\begin{eqnarray}
\label{5}
(nu^{\alpha})_{;\alpha}=0.
\end{eqnarray}
With Eq. (\ref{4}), we can rewrite (\ref{5}) in a differential form
\begin{eqnarray}
\label{6}
[(n/h)\psi^{,\alpha}]_{;\alpha}=0,
\end{eqnarray}
where $\psi^{,\alpha}$ means the contravariant component of the vector $\psi_{,\alpha}$. In general, Eq. (\ref{6}) is a nonlinear differential equation about $\psi$ and its derivatives. However, if $h$ is proportional to $n$, it will become a linear differential equation. Considering the simplest case, $P=\rho\propto n^{2}$, which implies that the speed of sound is equal to the speed of light and that the adiabatic index is equal to 2. The velocity of fluid must be subsonic everywhere, so no shock waves will arise. Then Eq. (\ref{6}) now reads
\begin{eqnarray}
\label{7}
(\psi^{,\alpha})_{;\alpha}=0.
\end{eqnarray}
We must solve this differential equation under appropriate boundary conditions to obtain physical quantities of accretion which we desire. One of the physical quantities of accretion we want to get is the particle number accretion rate
\begin{eqnarray}
\label{8}
\dot{N}=-\int_{S}nu^{i}\sqrt{-g}dS_{i}=-\int_{S}\psi_{,r}g^{rr}\sqrt{-g}d\Omega,
\end{eqnarray}
where $S$ represents the boundary two-dimensional sphere centered on the black hole, $r$ is the radial ordinate, and $g^{rr}$ is the radius-radius  component of the contravariant metric tensor, $g$ denotes the determinant of the metric of the black hole under considering, and $d\Omega$ is the product of the differentials of angels.

Here we consider a stationary accretion of an ultra-hard perfect fluid onto a charged dilaton black hole obtained in string theory \cite{Gibbons:1987ps, Garfinkle:1990qj}. The time and radial part of the metric of the charged dilaton black hole is exactly the same as that of the Schwarzschild black hole, but only the transverse parts of the metric are different, see Eq. (\ref{dil}), so it is asymptotically flat. We consider the first boundary condition, i.e., the asymptotic boundary condition at large distances. Since the medium is homogeneous there, we can take $n_{\infty}=h_{\infty}=1$ in appropriate units and restore $n_{\infty}$ in some final results. The asymptotic boundary condition in rectangular coordinates is
\begin{eqnarray}
\label{9}
\psi=u_{\mu} x^{\mu}=-u_{\infty}^{0}t+\textbf{u}_{\infty}\cdot\textbf{x}.
\end{eqnarray}
In spherical coordinates, it becomes
\begin{eqnarray}
\label{10}
\psi=-u_{\infty}^{0}t+u_{\infty}r[\cos\theta\cos\theta_{0}+\sin\theta\sin\theta_{0}\cos(\phi-\phi_{0})].
\end{eqnarray}
for $r\rightarrow\infty$. The asymptotic three-velocity vector, $v_{\infty}$, can point in an arbitrary direction ($\theta_0, \phi_0$). According to special relativity at infinity, there exists the following relation
\begin{eqnarray}
\label{11}
u^{\mu}_{\infty}=(u^{0}_{\infty}, \textbf{u}_{\infty})=(1-v^2_\infty)^{-1/2}(1, \textbf{v}_\infty).
\end{eqnarray}
Here we consider the accreted flow which moves into the black hole. Since the accreted flow is continuous at every point in space-time, and no disruption or infinity should exists in the flow, the physical quantities of the flow, such as the particle number density $n$ and the enthalpy $h$, should be finite everywhere, including at the event horizon.

\section{Accretion onto a moving charged dilaton black hole}
In this section, we consider accretion onto a charged dilaton black hole which moves at a constant velocity through the medium. In heterotic string theory, the dilaton has a linear coupling to the square of electromagnetic field tensor, so every solution with nonzero electromagnetic field tensor must have a nonconstant dilaton. Thus the Reissner-Nordstr\"{o}m solution, which describes charged black holes in general relativity, is not even an approximate solution in string theory. A solution for a statically charged black hole in the string theory was found to take the form \cite{Gibbons:1987ps, Garfinkle:1990qj}
\begin{eqnarray}
\label{dil}
ds^{2}=-\left(1-\frac{2M}{r}\right)dt^{2}+\left(1-\frac{2M}{r}\right)^{-1}dr^{2}+r(r-a)(d\theta^{2}+\sin^{2}\theta d\phi^{2}),
\end{eqnarray}
where $a=Q^{2}e^{2\phi_{0}}/M$ with $\phi_{0}$ the asymptotic constant value of the dilaton, $M$ the gravitational mass and $Q$ the charge of black hole, respectively. The event horizon are localized at $r =2M$. When $r=a$, the area of the sphere goes to zero and the surface is singular. The transition between black holes and naked singularities occurs at $Q=Q_{\rm{max}}\equiv \sqrt{2}e^{-\phi_{0}}M$. For $Q< Q_{\rm{max}}$, the singularity is enclosed by the event horizon. Since $g_{\rm{\theta\theta}}\geq0$ and $r> 2M$, we have a limit on parameter $a$: $a\leq 2M$. Considering the metric (\ref{dil}), Eq. (\ref{7}) implies
\begin{eqnarray}
\label{13}
-\frac{1}{1-\frac{2M}{r}}\frac{\partial^{2}\psi}{\partial t^{2}}+\frac{1}{r(r-a)}\frac{\partial}{\partial r}\left[r(r-a)\left(1-\frac{2M}{r}\right)\frac{\partial\psi}{\partial r}\right]
+\frac{1}{r(r-a)}\frac{1}{\sin\theta}\frac{\partial}{\partial\theta}\left(\sin\theta\frac{\partial\psi}{\partial\theta}\right)
+\frac{1}{r(r-a)\sin^{2}\theta}\frac{\partial^{2}\psi}{\partial\phi^{2}}=0.
\end{eqnarray}
 Taking into account the asymptotic boundary condition: $\psi=-u_{\infty}^{0}t+u_{\infty}r[\cos\theta\cos\theta_{0}+\sin\theta\sin\theta_{0}\cos(\phi-\phi_{0})]$  for $r\rightarrow\infty$, the general formula of $\psi$ can be assumed to take form $\psi=-u_{\infty}^{0}t+u(r,\theta,\phi)$, which also satisfies the stationary flow condition that the gradient of $\psi$ must be independent of time. Inserting $\psi$ in the Eq. (\ref{13}) with the general form, yields
\begin{eqnarray}
\label{14}
\frac{\partial}{\partial r}\left[r(r-a)\left(1-\frac{2M}{r}\right)\frac{\partial u}{\partial r}\right]+\frac{1}{\sin\theta}\frac{\partial}{\partial\theta}\left(\sin\theta\frac{\partial u}{\partial\theta}\right)
+\frac{1}{\sin^{2}\theta}\frac{\partial^{2} u}{\partial\phi^{2}}=0.
\end{eqnarray}
This is the differential equation which the spatial component $u(r,\theta,\phi)$ of $\psi$ should satisfy. If assuming $u=R(r)\Theta(\theta)\Phi(\phi)$, then the functions $R(r)$, $\Theta(\theta)$, and $\Phi(\phi)$, respectively, satisfies the differential equations below
\begin{eqnarray}
\label{15}
\frac{d^{2}\Phi}{d\phi^{2}}+m^{2}\Phi=0,
\end{eqnarray}
\begin{eqnarray}
\label{16}
\frac{1}{\sin\theta}\frac{d}{d\theta}\left(\sin\theta\frac{d\Theta}{d\theta}\right)-\frac{m^{2}\Theta}{\sin^{2}\theta}+l(l+1)\Theta=0,
\end{eqnarray}
and
\begin{eqnarray}
\label{17}
\frac{d}{dr}\left[r(r-a)\left(1-\frac{2M}{r}\right)\frac{d R}{dr}\right]-l(l+1)R=0.
\end{eqnarray}
Obviously, Eq. (\ref{17}) is complex and difficult to solve directly. The key here is to introduce a new variable to simplify it. After some attempts, we found a variable, $\xi=\frac{2}{2M-a}r-\frac{2M+a}{2M-a}$, to simplify Eq. (\ref{17}) as
\begin{eqnarray}
\label{18}
(1-\xi^{2})\frac{d^{2}R}{d\xi^{2}}-2\xi\frac{dR}{d\xi}+l(l+1)R=0,
\end{eqnarray}
which is Legendre equation. The general solutions for the differential equations (\ref{15}), (\ref{16}), and (\ref{18}) can take the following forms, respectively
\begin{eqnarray}
\label{19}
\Phi(\phi)=C\cos m\phi+D\sin m\phi~~~~m=0,1,2,\cdots,
\end{eqnarray}
where $C$, and $D$ are constants;
\begin{eqnarray}
\label{20}
\Theta(\theta)=P_{l}^{m}(cos\theta)~~~~l,m=0,1,2,\cdots,
\end{eqnarray}
where $P_{l}^{m}(z)$ is the associated Legendre function; and
\begin{eqnarray}
\label{21}
R(\xi)=AP_{l}(\xi)+BQ_{l}(\xi)~~~~l=0,1,2,\cdots,
\end{eqnarray}
where $A$ and $B$ are constants, $P_{l}(z)$ is the Legendre polynome, and $Q_{l}(z)$ is linearly independent from $P_{l}(z)$ and belongs to the second kind of Legendre function. Then the general formula of $\psi$ for charged dilaton black hole (\ref{dil}) is
\begin{eqnarray}
\label{22}
\psi=-u^{0}_{\infty}t+\sum_{l,m}[A_{lm}P_{l}(\xi)+B_{lm}Q_{l}(\xi)]Y_{lm}(\theta,\phi)~~~~~~l, m=0, 1, 2, \cdots,
\end{eqnarray}
where $A_{lm}$, $B_{lm}$ are constants needed to be determined from boundary conditions, $Y_{lm}(\theta,\phi)$ are spheric harmoics made up from $\Phi(\phi)$ and $\Theta(\theta)$. According to the boundary conditions, we can determine the specific form of Eq. (\ref{22}). Since the particle number density is finite at the event horizon, we first obtain the four-velocities as
\begin{eqnarray}
\label{23}
nu_{t}=-u^{0}_{\infty},
\end{eqnarray}
\begin{eqnarray}
\label{24}
nu_{r}=\frac{2}{2M-a}\sum_{l,m}\left[A_{lm}P_{l}^{'}(\xi)+B_{lm}Q_{l}^{'}(\xi)\right]Y_{lm}(\theta,\phi),
\end{eqnarray}
\begin{eqnarray}
\label{25}
nu_{\theta}=\sum_{l,m}\left[A_{lm}P_{l}(\xi)+B_{lm}Q_{l}(\xi)\right]\frac{\partial Y_{lm}(\theta,\phi)}{\partial\theta},
\end{eqnarray}
\begin{eqnarray}
\label{26}
nu_{\phi}=\sum_{l,m}\left[A_{lm}P_{l}(\xi)+B_{lm}Q_{l}(\xi)\right]\frac{\partial Y_{lm}(\theta,\phi)}{\partial\phi}.
\end{eqnarray}
where the prime denotes the derivative with respect to $\xi$. Then, substituting the corresponding quantities above in the normalization condition, we get
\begin{eqnarray}
\label{27}
n^{2}=\frac{(nu_{t})^{2}}{1-\frac{2M}{r}}-\left(1-\frac{2M}{r}\right)(nu_{r})^{2}-\frac{(nu_{\theta})^{2}}{r(r-a)}-\frac{(nu_{\phi})^{2}}{r(r-a)\sin^{2}\theta}.
\end{eqnarray}
Thinking of the limit behaviour of the Legendre functions and the finiteness of $n$ at the event horizon where $\xi=1$, we have
\begin{eqnarray}
\label{28}
n^{2}\rightarrow\frac{1}{1-\frac{2M}{r}}\left\{(u^{0}_{\infty})^{2}-\left[\frac{1}{4M}\sum_{l,m}B_{lm}Y_{lm}(\theta,\phi)\right]^{2}\right\}.
\end{eqnarray}
Observing the right hand side of the equation above, we find that the denominator in the external part will tend to zero when closing to the event horizon, the internal part must be equal zero at the same time to ensure that the particle number density is finite at the event horizon. Therefore we find
\begin{eqnarray}
\label{29}
B_{00}Y_{00}=4Mu^{0}_{\infty}
\end{eqnarray}
and all other $B_{lm}$s are zero. Therefore, Eq. (\ref{22}) reduces to
\begin{eqnarray}
\label{30}
\psi=-u^{0}_{\infty}t-2Mu^{0}_{\infty}\ln{\frac{r-2M}{r-a}}+\sum_{l,m}A_{lm}P_{l}(\xi)Y_{lm}(\theta,\phi),
\end{eqnarray}
where $A_{lm}$ can now be determined from the asymptotic boundary conditions in Eq. (\ref{11}). Without loss of generality, we focus on the case of $\theta=0$ which means the accretion flow at infinity moving towards the north pole of the coordinate system. Then the asymptotic boundary condition becomes
\begin{eqnarray}
\label{31}
\psi=-u^{0}_{\infty}t+u_{\infty}r\cos\theta.
\end{eqnarray}
Comparing it with Eq. (\ref{30}), yields
\begin{eqnarray}
\label{32}
A_{10}=\frac{2M-a}{2}u_{\infty},
\end{eqnarray}
and all other $A_{lm}$s are zero. Finally, we find the solution $\psi$ takes the form
\begin{eqnarray}
\label{33}
\psi=-u^{0}_{\infty}t-2Mu^{0}_{\infty}\ln{\frac{r-2M}{r-a}}+u_{\infty}\left(r-\frac{2M+a}{2}\right)\cos\theta.
\end{eqnarray}
Using Eq. (\ref{4}) and the final expression of $\psi$ in Eq. (\ref{33}), the four-velocity of accreted flow is calculated as
\begin{eqnarray}
\label{34}
nu_{t}=-u^{0}_{\infty},
\end{eqnarray}
\begin{eqnarray}
\label{35}
nu_{r}=-\frac{2Mu^{0}_{\infty}(2M-a)}{(r-a)(r-2M)}+u_{\infty}\cos\theta,
\end{eqnarray}
\begin{eqnarray}
\label{36}
nu_{\theta}=-u_{\infty}\left(r-\frac{2M+a}{2}\right)\sin\theta,
\end{eqnarray}
\begin{eqnarray}
\label{37}
nu_{\phi}=0.
\end{eqnarray}
The velocity field is a function of the coordinates $r$ and $\theta$, meaning that it possesses axial symmetry. The velocity also dependents on the parameter $a$ which is contained in the transverse parts of the metric. For radial accreted matter, it is generally believed that the transverse parts of the metric will not affect the velocity of the flow, but our results denies this view.

Substituting the components of four-velocity, Eqs. (\ref{34})-(\ref{37}), into Eq. (\ref{27}), we obtain the particle number density
\begin{eqnarray}
\label{38}
\nonumber
n^{2}&=&(u^{0}_{\infty})^{2}\left\{\frac{1}{1-\frac{2M}{r}}\left[1-\left[\frac{2M(2M-a)}{r(r-a)}\right]^{2}\right]\right\}
-(u_{\infty})^{2}\left\{\left(1-\frac{2M}{r}\right)\cos^{2}\theta+\frac{r}{r-a}
\left(1-\frac{2M+a}{2r}\right)^{2}\sin^{2}\theta\right\}\\
&+&(u^{0}_{\infty}u_{\infty})\frac{4M(2M-a)}{r(r-a)}\cos\theta.
\end{eqnarray}
From this equation, we can judge that the particle number destiny is really finite at event horizon. By solving the corresponding equations, we can determine that the stagnation point at which the velocity is zero lies at $\theta=0$ (directly downstream) and at radius
\begin{eqnarray}
\label{39}
r=\frac{1}{2}\left\{2M+a+\left[(2M+a)^{2}-8M\left(a+\frac{a-2M}{v_{\infty}}\right)\right]^{1/2}\right\}.
\end{eqnarray}
With Eq. (\ref{8}), we find the particle number accretion rate is given by
\begin{eqnarray}
\label{40}
\dot{N}=8\pi M(2M-a)u^{0}_{\infty}n_{\infty},
\end{eqnarray}
where we have restored $n_{\infty}$. Since $\dot{N}\geq 0$, meaning that $a\leq 2M$, which is consistent with the limit constrained from equation (\ref{dil}). The right hand side of Eq. (\ref{40}) is just the area of the black hole multiplied by particle number destiny and Lorentz factor for the accreted flow at infinity. Since $a$ depends on the magnetic charge and the dilaton, which means that these parameters play important parts in the process of accretion. If $a=0$, all the results obtained here reduce to the case of Schwarzschild black hole. According to the observations, we give a rough limit on the parameter $a$. Taking $c\sim 2.998\times 10^{10}$cms$^{-1}$, $G\sim 6.674\times 10^{-8}$cm$^{3}$g$^{-1}$s$^{-2}$, $M\sim 10M_{\odot}\sim 1.989\times 10^{34}$g, $m\sim m_{p}\sim 1.67\times 10^{-24}$g, $n_{\infty}\sim 1$cm$^{-3}$, $u^{0}_{\infty}\sim 2.996\times 10^{10}$cms$^{-1}$, and $\dot{M}\lesssim \dot{M}_{\rm{Schwarzschild}}$, we find $a \lesssim 2.6\times 10^{5}$cm.

\section{Conclusions and discussions}
We have obtained an analytic solution for accretion of a gaseous medium with a adiabatic equation of state $(P=\rho)$ onto a charged dilaton black hole which moves at a constant velocity. We have derived the four-velocity of accreted flow and found that it possesses axial symmetry. We have determined the particle number density and the location of the stagnation point. We have presented the accretion rate which depends on the the mass of black hole and the parameter $a$ in string theory. We also have discussed the possible theoretical and observational constraints on the parameter $a$. Since $a$ depends on the magnetic charge and the dilaton, which indicates that these parameters play important parts in the process of accretion onto the moving charged dilaton black hole. The time and radial part of the metric of the charged dilaton black hole we consider is exactly the same as that of the Schwarzschild black hole, only the transverse parts of the metric are different. However our solution is very different from that of Schwarzschild case. For radial accreted matter, it is generally believed that the transverse parts of the metric will not affect the velocity of the flow, but our results denies this taken for granted view. This is a interesting result which may help us to get deeper understanding of the behavior of accreted flow near the event horizon of black hole. For further studies, one can use Gamma ray burst or X-ray observations (see for example \cite{Kazempour:2022asl, Panotopoulos:2021ezt}) to constrain the parameter $a$.

\begin{acknowledgments}
This work is supported in part by Hebei Provincial Natural Science Foundation of China (Grant No. A2014201068 and A2021201034).
\end{acknowledgments}

\bibliographystyle{ieeetr}
\bibliography{acc}
\end{document}